# Hybrid Integration of Quantum Cascade Lasers with Germanium-on-Silicon waveguides for Mid-Infrared Sensing Applications

COLIN J. MITCHELL,[1,*] LONGQI ZHOU,[2] KE LI,[1] DANIEL ADEYEMI,[1] AHMED OSMAN,[1] MILOS NEDELJKOVIC,[1] GLENN CHURCHILL,[1] JAMES C. GATES,[1] GRAHAM T. REED,[1] KRISTIAN M. GROOM,[2] JON HEFFERNAN,[2] GORAN MASHANOVICH,[1]

[1] *Optoelectronics Research Centre, University of Southampton, University Road, Southampton, Hampshire SO17 1BJ, UK*
[2] *Department of Electronic and Electrical Engineering, Sir Frederick Mappin Building, University of Sheffield, Mappin Street, Sheffield, S1 3JD, UK*
*c.j.mitchell@soton.ac.uk*

**Abstract:** We present a novel scheme for hybrid integration of quantum cascade laser bars with germanium-on-silicon waveguides operating in the mid-infrared. The laser bars are flip-chip bonded onto a germanium-on-silicon target chip without active alignment, achieving end-fire coupling efficiencies of up to 45% (3.5 dB loss) in pulsed operation. Optical power estimates indicate 20–30 mW coupled into the waveguide. The passive alignment approach, combined with a CMOS-compatible photonic integrated circuit fabrication process, offers a scalable pathway to fully integrated mid-infrared photonic systems for sensing, free-space communications, and the realization of novel light sources.

## 1. Introduction

Mid-infrared (MIR) silicon photonics, has become increasingly important due to the molecular absorption lines found in this spectral region (2-20 µm), which allow for identification of specific molecules based on individual chemical bonds, hence the term 'fingerprint region'. Applications for widespread environmental monitoring, e.g. greenhouse gases and pollutants, security applications, and high-volume use of novel medical sensors for screening or monitoring of patients, require low-cost compact systems. Developing such MIR systems for use outside of a laboratory would benefit significantly from the advantages that silicon photonic integration [1] provides, namely small footprint, light weight, mass production and low cost [2, 3]. Integrating light sources onto the chips will be crucial for most of the aforementioned applications preferably with conceptually similar designs for all MIR wavelengths.

The indirect band gap of silicon prohibits the demonstration of lasers in the material, so the integration of compound semiconductor lasers has thus become a critical technology for the industry for the realization of compact practical devices with several competing approaches being proposed [4]. The majority of this work has been performed at near-infrared (NIR) communication wavelengths where techniques such as wafer bonding, transfer printing [5-7] and direct growth [8-10] have been demonstrated, along with flip-chip bonding of processed lasers[11]. However, laser integration within the MIR region of the spectrum has not been widely addressed. Quantum cascade lasers (QCLs) are commonly used as light sources in the MIR because QCL heterostructures can be designed to have intersubband transitions with energy levels that correspond to photon emission at MIR wavelengths [12]. However, integrating MIR lasers gives rise to some key additional challenges. High QCL power consumption creates thermal management issues, and the limited transparency range of $SiO_2$ prevents the use of widely used NIR SOI or SiN waveguides, therefore alternative mid-IR transparent waveguide platforms must be used instead (e.g. Ge-on-Si, SiGe-on-Si, or suspended

Si). At mid-IR wavelengths the optical mode sizes increase, therefore the dimensions of these waveguides must be significantly larger than NIR waveguides, which can simultaneously increase the tolerance for laser to waveguide alignment, along with increased mode overlap, and complicate the fabrication and integration process.

Integration of a QCL with silicon photonic waveguides was first demonstrated on a silicon substrate [13] at 4.7 µm, by heterogeneously bonding a QCL gain block to a silicon-on-nitride-on-insulator (SONOI) photonic integrated circuit (PIC) platform. The light was evanescently coupled between the QCL waveguide and the SONOI waveguide. The fabrication process involved several steps being performed on the hybrid III-V/Si stack after bonding, which limits the use of this integration method to foundries/tools in which contamination between III-V and Si materials is accepted. Output powers of around 31 mW were reported in pulsed operation from a coupled waveguide at 20ºC. Laser arrays of distributed feedback (DFB) and distributed Bragg-reflection (DBR) lasers were subsequently demonstrated and integrated with arrayed waveguide grating (AWG) multiplexers [14] on a silicon-on-insulator (SOI) platform.

The direct growth of QCLs on silicon substrates was first demonstrated with output power up to 50 mW per laser facet [15], emitting at a wavelength around 11 µm. Silicon (100) substrates 6º miscut towards [110] were utilised to reduce anti-phase domain [16] formation. Further development has included ICLs [17] at shorter wavelengths [18, 19] demonstrating CW output powers of 40 mW at room temperature at 3.3 µm. In the far infrared, an InP-based QCL grown on (001) silicon, 4º miscut toward [110], with an InP epitaxial template [20] has demonstrated up to 4 W of output power from the laser facets at room temperature. These reports do not demonstrate the integration of lasers with MIR waveguides and large miscut angles are not compatible with manufacturing of silicon photonic devices. Integration issues remain such as the induced roughness in the waveguide from the miscut angle that is beneficial for growth in localizing defects to the interface, and critically thermal budgets where growth may include a desorbing process for the native oxide around 1000ºC, well outside the allowable limits for in-situ silicon nitride or germanium waveguides. High temperatures also need to be carefully considered in regards the doping profiles required for circuit components. Critically for silicon photonics this may require front-end growth of compound semiconductors leaving substrates incompatible with CMOS facilities.

Recently, flip-chip placement of a DFB QCL at 7.2 µm has been demonstrated with peak CW power of 0.7 mW [21] coupled to a GOS waveguide. This system used a support structure and self-alignment by solder surface tension upon heating [22, 23] with the solder deposited by screen printing. The final position of the chip, and hence yield, is highly dependent upon several factors including: tolerances of volume of deposited material, fabrication issues for solder wetting of the bond pads and friction-force ratios limiting the size of the chip for high yields.

In this paper we present an alternative approach for integration using QCL bars to provide multiple sources simultaneously coupled to germanium-on-silicon substrates, providing a longer wavelength operation when compared to silicon nitride waveguides. The technique can therefore be applied across the MIR, whilst representing a CMOS compatible solution with a scalable route to production. Laser bars can also be used simultaneously as a detector in one part and as an emitter in an adjacent region. Additionally, multiple lasers could be combined in circuit for increased power, or combining multiple wavelengths [14] selected using DFB QCL arrays such as for free space absorption measurements [24]. The technique presented here uses no active alignment, instead relying on etched support structures for passive vertical alignment and the accuracy of the flip-chip bonder in the contact plane. Bonding alignment is provided directly from the facets of the QCL and GOS waveguide using high magnification optics eliminating potential fiducial-to-facet distance variation in fabrication of either component. Solder is evaporated and patterned using standard processing tools allowing full back-end integration of the QCLs entirely within cleanroom facilities reducing the risk of contamination induced misalignment. The development of this approach yielded the first hybrid integration of a QCL on a silicon photonics platform in 2022 [25, 26], but with only <1 mW of output power.

Our new approach results in up to 45% peak coupling efficiency between the QCL and waveguide, and a maximum 20-30mW of optical power in the waveguide in pulsed operation, which is crucial for sensitive mid-IR sensors. The wavelength utilized, 5.7 µm, is important for medical sensing applications such as therapeutic drug monitoring (TDM) of phenytoin in patients, which is a drug that is used to control various types of seizures [27-29], as well as for the determination of oestrogen levels [30] useful for the diagnosis of breast cancer [31]. Whilst the wavelength in question is relevant to specific clinical needs it is important to note that this approach demonstrates a transferable technique across the MIR.

## 2. Integration design

Hybrid flip-chip bonding of processed QCLs to GOS circuits allows for the manufacture and optimization of the two components at dedicated, separate sites for group IV and compound semiconductors. However, this requires a co-design approach incorporating the integration mechanics and related challenges. Fig. 1 shows schematics illustrating the relative positioning of the QCL bar, GOS waveguide, and contact pads to allow for alignment of the modes for end-fire coupling between the QCL and the GOS waveguide, whilst maintaining electrical contact. The GOS contact pad that connects to the QCL via a solder, extends outside of the QCL footprint to enable contacts using electrical probes or bond wires. The waveguide includes two 90° bends (0.2 mm radius) spatially separating the QCL input from the waveguide output, Fig. 1(b), to reduce the probability of emission from the laser facet interfering with the output measurement at the waveguide facet. Gold-tin was used for the solder, e-beam evaporated from an alloy source and patterned using conventional lithography and lift-off techniques. An excess of solder, larger than the designed gap (400 nm) between bond pads provided by the support structure, described later, allowed for variation of etch depth across the substrate of ~200 nm, whilst providing reliable Ohmic contacts. The gold-tin solder had a composition of 25 wt% tin, giving a eutectic melting point of around 290°C [32] and a soldering temperature of 320°C.

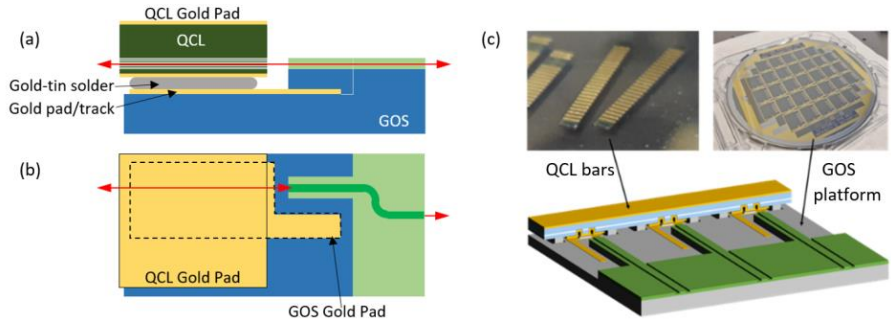

Fig. 1. (a) Side view of the QCL flip-chip bonded to the GOS chip showing the coupling region. (b) Top view of bonded QCL to GOS with the GOS gold pad shown as a dashed line. (c) QCL bars (top-left), processed GOS 6" wafer (top-right) and schematic of the bonded device (bottom).

To measure the coupling efficiency between the QCL and waveguide requires a measurement, or estimate, of the output power of the QCL towards the waveguide. As the QCLs are uncoated Fabry-Pérot (FP) lasers, the outputs from either facet are assumed to be equal, i.e. the power from the coupling facet to be equal to the uncoupled QCL back facet. Hence, we designed the chip such that light from the uncoupled QCL facet could be measured as an estimate of the power from the coupled QCL facet (Fig. 1). The GOS chips were diced using a self-polishing ductile dicing process [33, 34] which yields sufficient smoothness for measuring outputs without additional polishing. Standard loss measurement structures, varying the number of bends or length, were included on each die to assess the losses in the waveguide.

The GOS platform used in this work incorporates a 3-µm-thick germanium layer grown by Reduced Pressure Chemical Vapour Deposition (RPCVD) on (100) silicon [33]. Single mode GOS rib waveguides were designed for TM polarization, resulting in a width of 3 µm and a channel etch of 2.1 µm. The relatively thick waveguide layer, compared to SOI platforms, more closely matches the spot size of the laser and so alleviates the need for 3D spot size converters to achieve good modal overlap for end-fire coupling. To allow for some misalignment with improved modal overlap the waveguide is tapered up to 11 µm wide over a length of 1 mm. Additionally, the waveguide facet was angled at 10º to reduce reflection back into the QCL.

The chips were integrated by placing the QCL on top of the GOS with the grown surface face-down, using a FineTech Fineplacer® Lambda flip-chip bonder with an alignment accuracy ≤ 0.5 µm in the plane of the interface. Automated bonders with higher precision are available, demonstrating a route for scaling the process. Vertical and angular alignment out of this plane is not as accurate, and so the design of the components requires a way to address this issue as an initial consideration. Etched support structures in the GOS were chosen to passively align the QCL to the waveguide plane using the unprocessed QCL top epitaxial layer as the other contact to minimize error from etch or deposition on the QCL. The etched support structure can be seen in Fig. 1(c), but is not shown in Fig. 1 (a/b) for schematic clarity. Parallel placement, where the GOS supports are in contact with the QCL along the length of the laser bar (14.5 mm by 2.0 mm bar, with 24 lasers along the length at a pitch of 500 µm), requires the use of a dedicated gimballed placement head and other tooling described elsewhere [35].

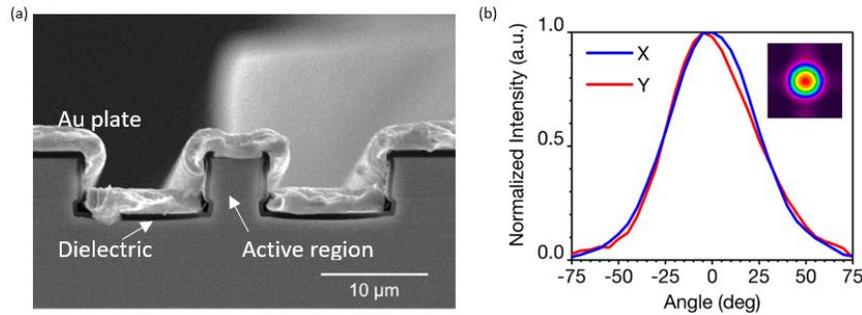

Fig. 2. (a) SEM image of a cross section in a 4.5µm wide double-trench QCL. (b) Measured X&Y-axis far-field beam profile, inset 2D colour plot (measured using an Ophir Nano Scan pyroelectric scanning-slit beam profiler).

The MIR QCLs are based on a two-phonon resonance design [36] using the strain-balanced InGaAs/AlInAs material system grown by metalorganic vapour-phase epitaxy (MOVPE) at the National Epitaxy Facility, Sheffield. The material was processed into double-trench ridge waveguide structures with ridge width of ~4.5 µm. Trenches penetrate through the active region to a depth of 6 µm, as shown in Fig. 2(a). The double trench scheme allowed the unprocessed epitaxial surface in between the laser waveguide and metal to be in contact with the support structures on the GOS substrate. Uncoated FP lasers mounted on diamond heat spreaders with the epitaxial surface facing up exhibit room temperature continuous-wave (RT-CW) operation, with a threshold current density of 0.963 kAcm$^{-2}$ for a 3 mm long cavity, and maximum power >65 mW per facet at a heatsink temperature of 15ºC. The measured optical far-field beam profile shown in Fig. 2(b) demonstrates a highly circular fundamental transverse mode, providing good beam quality, which is beneficial for effective waveguide coupling without 3D tapers. The FWHM fast- and slow-axis beam divergences are both around 55º.

### 3. Coupling measurements

Initial characterization of the integrated device demonstrated consistent electrical connections. In pulsed mode operation a far-field beam profile was measured from both a GOS waveguide facet (Fig. 3) and a grating coupler. To quantify the outputs from the device, measurements were made at QCL and GOS waveguide facets (Fig. 4), with light collected and collimated using a NA=0.85 lens (Thorlabs C037TME-E) with anti-reflection coating and focused on to a thermoelectrically cooled (TEC) Vigo InAsSb superlattice detector (PVIA-4TE-10.6-1x1) via a lens (Thorlabs LA7477-E2) with NA selected to match the detector acceptance angle. In the collimated section between the lenses, Thorlabs neutral density filters (NDF) were used to prevent saturation of the detector. The attenuation of the NDFs with nominal OD = 0.3, 1.0, 2.0, were experimentally determined using an identical QCL as the source.

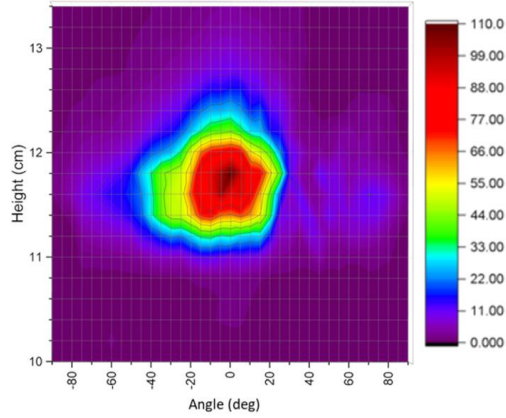

Fig. 3 Far-field image of light output from waveguide facet measured with a VIGO PVI-4TE-8 MCT-detector on a rotating stage, with height adjustment, at a distance of ~10cm

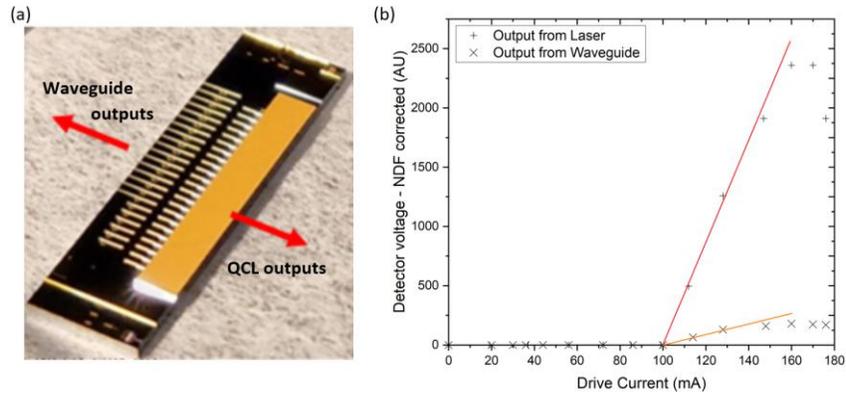

Fig. 4. (a) Integrated chip showing measured output facets. (b) QCL and GOS waveguide outputs from the integrated chip.

The laser to waveguide optical coupling was calculated using Eq. (1), shown schematically in Fig. 5. Fresnel reflection loss ($L_r$ = 1.89 dB) was calculated using the effective index ($n_{eff}$ = 3.93) determined from simulation of the taper structure. Propagation and bend losses ($L_p$ and $L_b$ respectively) were derived from the passive test structures using the cut-back method. The propagation loss was calculated as 8.6 dB/cm and the bend loss 9.8 dB/cm (0.38 dB/90° bend). The high loss through the waveguide is thought to result from the presence of strain-induced defects at the silicon-germanium interface, accentuated by the thermal cycling involved in the deep ICP-RIE etch. This will be improved in future devices by the use of cooler etch processes, which have been developed.

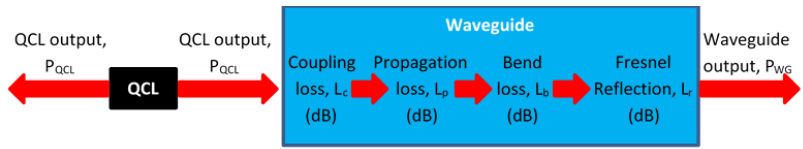

Fig. 5. Laser to waveguide optical coupling with losses.

$$L_c = -10 \cdot \log_{10}\left(\frac{P_{WG}}{P_{QCL}}\right) - L_p - L_b - L_r \qquad (1)$$

The data derived through this analysis is shown in Fig. 6 for the device with the highest coupling shown for two different pulse widths. The peak output coupling (output from integrated waveguide as a fraction of QCL output power) was 12.9% (8.9 dB loss). This produces a maximum value for laser to waveguide optical coupling of 45% (3.5 dB loss). As voltage rollover of the laser output begins (i.e. at drive current ~170 mA), the coupling falls to 26%. This reduction could be indicative of heating effects in terms of expansion or spectral changes in the FP cavity. The measurement was repeated with longer pulses (600 ns) and this provided more stable interface coupling values with increasing current, but with coupling limited to 20-27%. These values are similar to the values at 400 ns with higher current. This further suggests that the fall in efficiency for the shorter pulse could be due to heating effects with increased drive current, matching those for longer pulses as the laser reaches a steady thermal equilibrium. Of note here is that as the QCL reached roll-over, the output power profile on the oscilloscope changed. The intensity peak occurred earlier in the pulse, with a subsequent fall before the end of the pulse. This would occur with excess heating during the pulse.

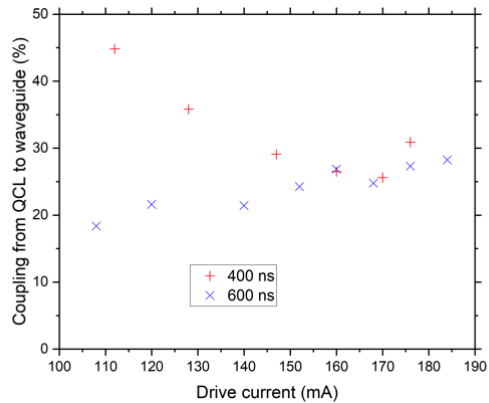

Fig. 6. Coupling efficiency against current for pulse widths of 400 ns and 600ns.

The data shown in Fig. 6 is from the QCL with the highest coupling efficiency on the bar. As the eleventh laser of twenty-four along the bar it was located close to the centre. Along the bar were a mixture of end-fire and grating outcoupled lasers. Seventeen QCLs were coupled to GOS waveguides with end-fire outcoupling, six QCLs were coupled to GOS waveguides with grating couplers at the output, and by design one had no waveguide interface. Grating couplers were not used for the calculation of the coupling between QCL and waveguide, due to a larger uncertainty in calculating the coupling efficiency by separating out grating coupler and fibre losses. Comparing the seventeen end-fire QCLs, one failed under initial electrical test, but all coupled light. Away from the centre of the bar, and in both directions, the coupling efficiency fell, indicating that this was not from rotational misalignment (yaw or roll). The fall in coupling efficiency, down to 3.5% at the edge of the laser bar (pulsed for 600 ns), is most likely due to the thermal expansion coefficient mismatch of the components at bonding (320°C). The potentially complex expansion mismatch (a function of tooling, components, and thermal gradients) can be experimentally measured and be repeatable. Incorporating this expansion into the design of the waveguide spacing can then increase the efficiency along the bar.

In order to obtain an estimate of the optical power in terms of mW, as opposed to detector voltages above, an identical QCL in a conventional laser driver unit was used for collaboration. Firstly, the power was measured in CW mode and then compared to the values of detector voltage, using a standard chopper (simulating pulsed measurement). Quantifying the output of the QCL and using the coupling efficiency at peak power, this equates to 20-30 mW of power in the waveguide, demonstrating the potential use of this system for sensing applications.

## 4. Conclusions

In this paper, we have demonstrated a novel integration of QCL laser bars to a silicon photonics platform using CMOS-compatible processes. The efficient co-design of the components opens up the possibility of scaling up production for widespread use of this system for applications such as medical, environmental monitoring and security sensing across the MIR.

Flip-chip bonding eliminates the issues from the material interface associated with die/wafer bonded and monolithically grown structures, whilst the gold-tin alloy solder is an industrial standard for contacts and acts as a critical heatsink for the QCL. Importantly, this approach allows for separate manufacture of the GOS and QCL components using optimal fabrication processes for each component with minimal back-end processing. Our hybrid approach using passive alignment of components, where the GOS contacts to unprocessed sections of the QCL, has demonstrated coupling into 11 µm wide waveguide facets along the length of a 14.5 mm laser bar. QCL performance has been demonstrated for pulsed operation after integration and demonstrating a coupling efficiency of up to 45%. Further improvements could be made by applying additional heatsinking to the exposed substrate of the QCL, proven to dramatically improve performance [11], and possibly enable CW and high power operation. Additionally, active-cooling of the QCL to circuits on the GOS could be utilized if required.

A drop off in coupling efficiency was noted away from the middle of the laser bar. This was expected and it is proposed that this is due to the mismatch in thermal expansion coefficients. A coupling efficiency between the laser and waveguide of 3.5% was still achieved close to the edge, representing an estimated ~1 mW in the waveguide.

These developments open the path to multiple sources, and the possibility of combining lasers at different wavelengths within the gain spectra of the QCLs, or combining detectors, all with a single alignment. With further developments such as active cooling for the QCL integrated on the GOS chip, portable low-cost, low power light sources on chip could be realized and become a common tool for tackling climate change and medical diagnosis.

**Funding.** The authors gratefully acknowledge funding from the Engineering and Physical Sciences Research Council (EP/V047663/1, EP/W035995/1, EP/N00762X/1, EP/W020254/1) and the Royal Academy of Engineering (RF201617/16/33).

**Datasets.** Data underlying the results presented in this paper are available with open access at: https://doi.org/10.5258/SOTON/D3466

**Disclosures.** The authors declare no conflicts of interest.

**References**

1. C. J. Mitchell, et al., "Mid-infrared silicon photonics: From benchtop to real-world applications," APL Photonics **9**, 080901 (2024).
2. R. Soref, "Mid-infrared photonics in silicon and germanium," Nature Photonics **4**, 495-497 (2010).
3. S. Shekhar, et al., "Roadmapping the next generation of silicon photonics," Nature Communications **15**, 751 (2024).
4. Z. Zhou, et al., "Prospects and applications of on-chip lasers," eLight **3**, 1 (2023).
5. G. H. Duan, et al., "New Advances on Heterogeneous Integration of III–V on Silicon," Journal of Lightwave Technology **33**, 976-983 (2015).
6. G. Roelkens, et al., "Micro-Transfer Printing for Heterogeneous Si Photonic Integrated Circuits," IEEE Journal of Selected Topics in Quantum Electronics **29**, 1-14 (2023).
7. M. A. Tran, et al., "Ring-Resonator Based Widely-Tunable Narrow-Linewidth Si/InP Integrated Lasers," IEEE Journal of Selected Topics in Quantum Electronics **26**, 1-14 (2020).
8. Z. Wang, et al., "Room-temperature InP distributed feedback laser array directly grown on silicon," Nature Photonics **9**, 837-842 (2015).
9. Y. Xue, et al., "In-Plane 1.5 µm Distributed Feedback Lasers Selectively Grown on (001) SOI," Laser & Photonics Reviews **18**, 2300549 (2024).

10. Y. De Koninck, et al., "GaAs nano-ridge laser diodes fully fabricated in a 300-mm CMOS pilot line," Nature **637**, 63-69 (2025).
11. D. Coenen, et al., "Thermal Characterisation of Hybrid, Flip-Chip InP-Si DFB Lasers," in *Micromachines,* (2023).
12. J. Faist, et al., "Quantum Cascade Laser," Science **264**, 553-556 (1994).
13. A. Spott, et al., "Quantum cascade laser on silicon," Optica **3**, 545-551 (2016).
14. E. J. Stanton, et al., "Multi-Spectral Quantum Cascade Lasers on Silicon With Integrated Multiplexers," Photonics **6**, 6 (2019).
15. H. Nguyen-Van, et al., "Quantum cascade lasers grown on silicon," Scientific Reports **8**, 7206 (2018).
16. H. Kroemer, "Polar-on-nonpolar epitaxy," Journal of Crystal Growth **81**, 193-204 (1987).
17. R. Q. Yang, "Infrared laser based on intersubband transitions in quantum wells," Superlattices and Microstructures **17**, 77-83 (1995).
18. A. Malik, et al., "Integration of Mid-Infrared Light Sources on Silicon-Based Waveguide Platforms in 3.5–4.7 μm Wavelength Range," IEEE Journal of Selected Topics in Quantum Electronics **25**, 1-9 (2019).
19. M. Fagot, et al., "Interband cascade lasers grown simultaneously on GaSb, GaAs and Si substrates," Optics Express **32**, 11057-11064 (2024).
20. S. Slivken and M. Razeghi, "High Power, Room Temperature InP-Based Quantum Cascade Laser Grown on Si," IEEE Journal of Quantum Electronics **58**, 1-6 (2022).
21. D. Wang, et al., "Innovative Integration of Dual Quantum Cascade Lasers on Silicon Photonics Platform," Micromachines **15**, 1055 (2024).
22. Y. Martin, et al., "Novel Solder Pads for Self-Aligned Flip-Chip Assembly," in *2019 IEEE 69th Electronic Components and Technology Conference (ECTC)*, 2019), 528-534.
23. Y. Martin, et al., "Flip-Chip III-V-to-Silicon Photonics Interfaces for Optical Sensor," in *2019 IEEE 69th Electronic Components and Technology Conference (ECTC)*, 2019), 1060-1066.
24. B. G. Lee, et al., "DFB Quantum Cascade Laser Arrays," IEEE Journal of Quantum Electronics **45**, 554-565 (2009).
25. C. J. Mitchell, et al., "Hybrid integration methodology for quantum cascade lasers with germanium waveguides in mid-IR," EPJ Web Conf. **266**, 01008 (2022).
26. C. J. Mitchell, et al., *Hybrid laser integration in the mid-IR for silicon photonics sensing applications*, SPIE OPTO (SPIE, 2023), Vol. 12426.
27. S. H. Jang, et al., "Therapeutic drug monitoring: A patient management tool for precision medicine," Clin Pharmacol Ther **99**, 148-150 (2016).
28. C. Hiemke, et al., "Consensus Guidelines for Therapeutic Drug Monitoring in Neuropsychopharmacology: Update 2017," Pharmacopsychiatry **51**, 9-62 (2018).
29. E. Kozer, et al., "How high can we go with phenytoin?," Ther Drug Monit **24**, 386-389 (2002).
30. B. Zheng, et al., "Separation and determination of estrogen in the water environment by high performance liquid chromatography-fourier transform infrared spectroscopy," Scientific Reports **6**, 32264 (2016).
31. Z. Diamantopoulou, et al., "The metastatic spread of breast cancer accelerates during sleep," Nature **607**, 156-162 (2022).
32. J. W. Ronnie Teo, et al., "Microstructure of eutectic 80Au/20Sn solder joint in laser diode package," Microelectronic Engineering **85**, 512-517 (2008).
33. M. Nedeljkovic, et al., "Germanium-on-silicon waveguides operating at mid-infrared wavelengths up to 8.5 μm," Optics Express **25**, 27431-27441 (2017).
34. P. C. Gow, et al., "Mechanical dicing of optical quality facets and waveguides in a silicon nitride platform," Electronics Letters **60**, e13138 (2024).
35. C. J. Mitchell, et al., "Tooling and procedures for hybrid integration of lasers by flip-chip technology," in *2020 IEEE 8th Electronics System-Integration Technology Conference (ESTC)*, 2020), 1-7.
36. I. I. Zasavitskii, et al., "A GaInAs/AlInAs quantum cascade laser with an emission wavelength of 5.6 μm," Quantum Electronics **48**, 472 (2018).